\documentstyle[prl,aps,epsfig,twocolumn]{revtex}
\begin{document}
\draft
\flushbottom
\twocolumn[
\hsize\textwidth\columnwidth\hsize\csname @twocolumnfalse\endcsname

\title{Consistency of superconducting correlations with one-dimensional
electron interactions in carbon nanotubes}
\author{J. Gonz\'alez   \\}
\address{
        Instituto de Estructura de la Materia.
        Consejo Superior de Investigaciones Cient{\'\i}ficas.
        Serrano 123, 28006 Madrid. Spain.}

\date{\today}
\maketitle
\begin{abstract}
\widetext
We show that a model of interacting electrons in one dimension
is able to explain the order of magnitude as well as the
temperature dependence of the critical supercurrents recently
measured in nanotube samples placed between superconducting
contacts. We use bosonization methods to compute superconducting
correlations in the presence of the long-range Coulomb
interaction, ending up with a picture in which the critical current
does not follow the temperature dependence of the gap
in the contacts, in contrast to the prediction of the conventional
proximity effect. 
Our results also reveal the presence of a
short-range attractive interaction in the nanotubes,
which accounts for a significant enhancement of the critical
supercurrents.

\end{abstract}

]

\narrowtext 
\tightenlines


Since the discovery of carbon nanotubes, these have offered a
great potential for novel electronic properties and 
technological applications. 
It has been checked experimentally\cite{wild}
the theoretical prediction that 
there should be semiconducting as well as metallic 
nanotubes\cite{ham}. 
It has been also remarkable the experimental observation of 
unconventional transport properties\cite{exp}, that seem to be 
compatible with the expected Luttinger liquid behavior of 
one-dimensional electron systems\cite{lutt}. 
Different approaches have predicted the 
appearance of phases with broken symmetry in the carbon nanotubes 
at very low energies\cite{bal,louie,eg,odint}. Anyhow, the estimates 
are in general that there should be enough margin to observe the 
characteristic scaling behavior of the Luttinger liquid over a wide
range of temperatures.

A different class of experiments has been aimed to test the
superconducting properties of the carbon nanotubes\cite{kas,marcus}. 
One of the most striking results has been the observation of 
supercurrents along carbon nanotubes placed between superconducting 
contacts\cite{kas}. In a sample made of a single-walled nanotube, 
for instance, critical supercurrents have been measured that are 
about 40 times higher than expected from the value of the gap in the
contacts\cite{kas}. They also show a very flat
dependence with temperature, until the critical value of the
superconducting contacts is approached. In that respect, there is a
marked difference from the behavior of another sample made of a rope  
of nanotubes, where the critical supercurrent seems to follow the
BCS gap in a certain range of low temperature\cite{kas}.

A model of the electron interaction in the carbon
nanotubes should give a quantitative account of all these
different observations of superconducting correlations. We show
in this letter that the mentioned features of the critical
supercurrent can be understood in the framework of a
one-dimensional theory of interacting electrons. We will see
that the experimental data are consistent with a definite
form of the one-dimensional interaction, as it is actually
nontrivial to reproduce both the shape and the order of
magnitude of the supercurrents in the single-walled nanotube and
in the rope of nanotubes. In particular, the experimental values 
of the supercurrent point at a sensible renormalization of the 
strength of the long-range Coulomb interaction, specially 
in the sample made of a rope of nanotubes, in agreement with 
earlier theoretical predictions\cite{eur,graph}.

A metallic single-walled nanotube has several one-dimensional
subbands, with two pairs of linear branches crossing at
Fermi points $k_F$ and $-k_F$.
We deal in this letter with an effective description of the
nanotubes for energies below the scale $E_c$ at which all the
gapped subbands decouple in the computation of low-energy
properties, so that the relevant modes left belong to the
linear branches close to the Fermi level. We
can estimate this energy $E_c$ as a few 
tenths of ${\rm eV}$, for a typical single-walled nanotube 
with about 10 subbands. 

The low-energy excitations can be encoded into four boson fields,
each boson corresponding to a linear branch in the same fashion
as in the Luttinger model\cite{lutt}. The hamiltonian of the 
effective theory
can be written in terms of the respective density operators 
$\rho_{i a \sigma}$, labelled by the Fermi point $a = 1,2$ and
by the chirality $i = L,R$,
\begin{eqnarray}
H  & = &      \frac{1}{2} v_F \int_{-k_c}^{k_c} dk 
 \sum_{i a \sigma }  : \rho_{i a \sigma} (k)
             \rho_{i a \sigma} (-k)  :   \nonumber      \\
  &  &    + \frac{1}{2}  \int_{-k_c}^{k_c} dk \; 
      \sum_{i a \sigma } \rho_{i a \sigma} (k) \; V(k)
      \sum_{j b \sigma'  } \rho_{j b \sigma'} (-k)
\label{ham}
\end{eqnarray}
In the above expression, $k_c$ is related to $E_c$ through the 
Fermi velocity $v_F$, $k_c = E_c/v_F$. 

Our assumption regarding the
interaction will be the presence of the long-range 
Coulomb interaction $V(k) \approx e^2 /(4\pi^2) \log |k_c / k|$, 
which remains unscreened in one spatial dimension\cite{grab}, 
plus an additional short-range effective attraction
coming from the coupling to the elastic modes of the nanotube. 
In this framework, we are neglecting backscattering and Umklapp
processes that mix different chiralities and Fermi points, relying on
the fact that those interactions have smaller relative strength
($\sim 0.1/n$, in terms of the number $n$ of subbands\cite{eg,kane})
and they remain small down to extremely low energies\cite{eg}.

The correlators in the model governed by (\ref{ham}) can be
computed by changing variables to the total charge density
operators 
\begin{equation}
\rho_i (k) = \frac{1}{\sqrt{N}} \sum_{a \sigma } 
                  \rho_{i a \sigma} (k)  \;\;\;\;\;\;  i = L,R
\end{equation}
where $N$ stands in general for the number of channels 
$\{ (a,\sigma) \}$, so that $N = 4$ in the case of a
single-walled nanotube.

A typical propagator of Cooper pairs, for instance, becomes
\begin{eqnarray}
 G(x,t) & \equiv &    \langle \Psi_{L 1 \uparrow} (x,t) 
   \Psi_{R 2 \downarrow} (x,t) \Psi^{+}_{L 1 \uparrow} (0,0)   
     \Psi^{+}_{R 2 \downarrow} (0,0)  \rangle  \nonumber  \\ 
      &  =  &   C(x,t) F(x,t)
\end{eqnarray}
where $F$ is the part that does not depend on the interaction and
$C$ corresponds to the propagation of the total charge. At zero
temperature, for instance, we have 
\begin{equation}
C(x,t) = \exp \left( -\frac{2}{N} \int_{0}^{k_c} dk \frac{1}
   {\mu (k)\: k} \left(1 - \cos (kx) \: \cos (\tilde{v}_F kt) \right)
            \right)
\label{prop}
\end{equation}
where $\mu (k) = 1/\sqrt{1 + 2N V(k)/v_F}$ and $\tilde{v}_F = 
v_F/\mu (k)$. The other factor has the simple dependence
\begin{equation}
F(x,t) = 1/ \left|k_c^2 (x - v_F t) (x + v_F t)\right|^{\frac{N-1}{N}}
\end{equation}

The critical supercurrent ${\cal I}$ can be estimated under
the assumption that (i) the normal-superconductor junctions are 
perfectly transmitting\cite{maslov}, or 
(ii) the single-particle scattering is
relevant at the interfaces\cite{fazio}. The latter is
more realistic for the experiments that we are considering. The
distance $L$ between the superconducting contacts is large
enough that ${\cal I}$ can be expressed as a function of $L$ and 
the temperature $T$ as 
\begin{equation}
{\cal I}_L (T)  = e v_F k_c \int_{0}^{1/T} d \tau \;  G(L, -i\tau )
\label{curr}
\end{equation}
In the above equation, $G$ stands
for the appropriate expression at finite temperature. The analytic
continuation to imaginary time, however, cannot be taken directly
in expressions like (\ref{prop}), and for computational purposes
it is more convenient to introduce the temperature dependence through 
the Matsubara formalism
\begin{eqnarray}
C(x,-i\tau) & = &  \exp \left( -\frac{2}{N} \int_{0}^{k_c} dk \;
   \frac{2T}{v_F}   \right.     \nonumber            \\  
    &  &  \left.        \sum_{m = -\infty}^{m = +\infty}
    \frac{1 - \cos (kx) \: \cos (2\pi m T \tau )}
   {(2\pi m T/\tilde{v}_F)^2 + k^2}  \right)
\end{eqnarray}

We can use Eq. (\ref{curr}) to test whether the behavior of the 
critical currents measured in Ref. \onlinecite{kas} can be 
reproduced within the present framework. The comparison should be
fairly direct for the sample that is made at one end of a single
nanotube (called $ST_1$ in Ref. \onlinecite{kas}). 
According to the above discussion, we consider a 
momentum-dependent parameter
$\mu (k) = 1/\sqrt{ 1 + N \left( e^2/(2\pi^2 v_F) \log |k_c /k|
 - g/(\pi v_F) \right) }$, taking in this case
a number of channels $N = 4$ in the above equations.

\begin{figure}[H]

\epsfig{file=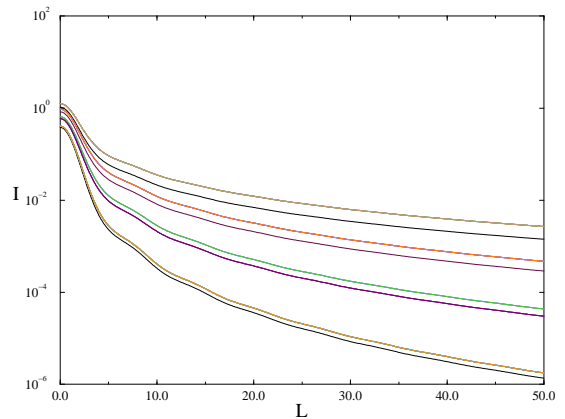, height=7cm, width=7cm,
         bbllx=0, bblly=0, bburx=600, bbury=600,
         angle=270}

\caption{Plots of the critical current (in units of $e v_F k_c$)
versus distance, at $T = 0$, for different strengths of the Coulomb
interaction. From top to bottom, the solid curves correspond to
$2 e^2/(\pi^2 v_F) = 1.0, 2.0, 4.0, 8.0$. The dotted (colored) lines
correspond in each case to the correction by effect of the additional
short-range interaction, with $4g/(\pi v_F) = 0.75$.}
\label{two}
\end{figure}

We have checked first that the critical current ${\cal I}
\approx 0.1 {\rm \mu A}$ of the $ST_1$ sample at $T \approx
0 {\rm K}$, found anomalously high in the BCS framework, 
can be explained with the present 
model. The results represented in Fig. \ref{two} show
the magnitude of ${\cal I}_L (0)$ for different values of the
interaction, the distance being measured
in units of $k_c^{-1}$. The actual values of the 
supercurrent are obtained by multiplying the magnitudes in Fig.
\ref{two} by the prefactor in Eq. (\ref{curr}).
The Fermi velocity can be obtained from the hopping amplitude
$t \approx 2.1 {\rm eV}$ and the nearest-neighbor distance
$a \approx 1.4 {\rm \AA} $, by using the expression $v_F = 3ta/2$.
A reasonable estimate of the cutoff $k_c$ for the single-walled
nanotube is $k_c \approx 0.5 {\rm nm}^{-1}$, which gives
$e v_F k_c \approx 30 {\rm \mu A}$.

We observe that the coupling corresponding to the bare parameters 
of a graphite layer, $2 e^2/(\pi^2 v_F) \approx 8.0$, does not lead 
to sensible results. The total length of the sample $ST_1$ is
$\approx 300 {\rm nm}$, and we should expect to get the correct
order of magnitude of the critical current at $L \sim 50/k_c$.
The most appropriate value for the Coulomb interaction seems to be 
$2 e^2/(\pi^2 v_F) \approx 1.0$. The order of magnitude ${\cal I} 
\approx 0.1 {\rm \mu A}$ is then reached, most precisely if one  
takes into account a coupling for the short-range attractive 
interaction $g/(\pi v_F) \sim 0.2$. The sensible reduction in 
the value of $2 e^2/(\pi^2 v_F)$ can be understood by the presence
of nearby charges and the renormalization of the 
interaction in the narrow rope into which the
nanotube merges, as we discuss afterwards.

\begin{figure}[H]

\epsfig{file=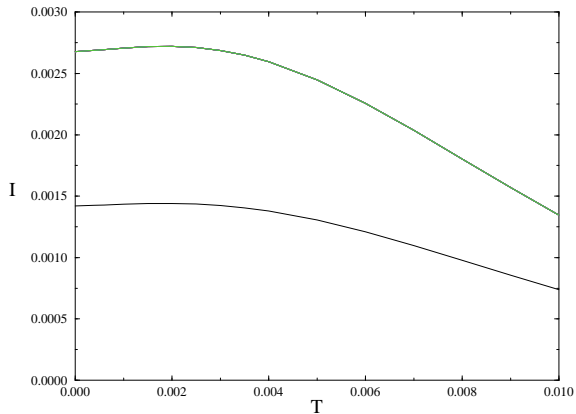, height=7cm, width=7cm,
         bbllx=0, bblly=0, bburx=600, bbury=600,
         angle=270}

\caption{Plots of the critical current (in units of $e v_F k_c$)
versus $T/E_c$, for $2 e^2/(\pi^2 v_F) = 1.0$ and a coupling of   
the short-range attractive interaction $4g/(\pi v_F) = 0$ (lower
curve) and 0.75 (upper curve).}
\label{three}
\end{figure}

Moreover, the dependence of the critical current on $T$ for the
mentioned interaction strength reproduces the shape that has been
observed in the measurements of the sample $ST_1$. In the 
theoretical model, the temperature $T$ is given in units of the 
unique energy scale $E_c$. This means that the critical temperature
$T_c \approx 0.4 {\rm K}$ of the contacts for the sample $ST_1$ 
corresponds to the 
dimensionless value $T_c/E_c \approx 2 \times 10^{-4}$.
We have represented in Fig. \ref{three} the results for the 
critical current ${\cal I}_L (T)$ at $L = 50/k_c$, with and
without the effect of the short-range attractive interaction. It is
remarkable the smooth behavior of the critical current in the
low-temperature regime below $T_c$. The slight increase observed
near zero temperature is related to the renormalization of the
transmission at the interfaces\cite{kf}. Near $T_c$, the suppression
of superconductivity in the contacts should be incorporated to
produce a sharp decrease, leading then to the full agreement with
the experimental results of Ref. \onlinecite{kas}.

Moving now to the sample made of a rope of nanotubes (called 
$R0_3$ in Ref. \onlinecite{kas}), we have to take into account
two main differences with respect to the preceding discussion.
First, the energy scale $E_c$ up to which the rope can be seen as a 
purely one-dimensional system is smaller, compared to the cutoff
introduced for the sample $ST_1$. Given that the diameter
of $R0_3$ can be approximately 15 times larger than that of 
the single-walled nanotube of $ST_1$, we may assume that the 
energy cutoffs in the two samples differ in the inverse proportion 
by a factor of 15. This means that a given temperature of
the sample $R0_3$, when measured in units of the corresponding 
$E_c$, looks comparatively higher than the same temperature in
the sample $ST_1$. Thus, the critical temperature $T_c \approx 
1.1 {\rm K}$ of the contacts used for the sample $R0_3$ gives 
a dimensionless value $T_c/E_c \approx 9 \times 10^{-3}$, 
which is more than one order of
magnitude higher than the ratio for the sample $ST_1$.

The second important difference between the rope of nanotubes 
$R0_3$ and the sample $ST_1$ is the interaction among the large 
number of nanotubes ($\approx 200$) in the former\cite{mele}. 
The interaction among the charge in the different channels produces
a significant renormalization of the strength of the Coulomb 
interaction. This can be understood
in the bosonization approach developed above, if we consider that
each metallic nanotube in the rope contributes with four units to
the number $N$ of channels that can be bosonized\cite{egger}.
A large number $N$ implies that
the contribution of the factor $C(x,t)$ to the supercurrent is
greatly diminished, which has in practice the same effect as 
reducing the strength of the interaction.
The picture is more involved considering the whole number of 
nanotubes in the rope, as the interaction with the charge in the 
semiconducting tubules cannot be completely neglected. The overall
physical effect can be taken into account by assuming a
scale-dependent renormalization of the coupling constant 
$2 e^2/(\pi^2 v_F)$ in the rope from its bare value at the scale of 
a few angstroms, as discussed in Ref. \onlinecite{eur}.  

Given that the total length of the sample $R0_3$ is 
$\approx 1.7 {\rm \mu m}$, we have 
estimated the supercurrent by the decay of ${\cal I}_L$ through a 
distance $L = 50/k_c \approx 1.5 {\rm \mu m}$. The evaluation of the 
prefactor in front of Eq. (\ref{curr}) is now more delicate, compared
to that for the sample $ST_1$. On the one hand, we have to 
bear in mind that the value of $k_c$ decreases according
to the increase in the diameter of the sample. On the other hand, 
there are more metallic nanotubes in the sample $R0_3$, in a number 
that may be estimated as $1/3$ of the 
total number, which gives $\approx 60$ metallic nanotubes. Balancing 
both points, it is appropriate to take now a prefactor in Eq. (\ref{curr}) 
that is four times the value for the single-walled nanotube. 

\begin{figure}[H]

\epsfig{file=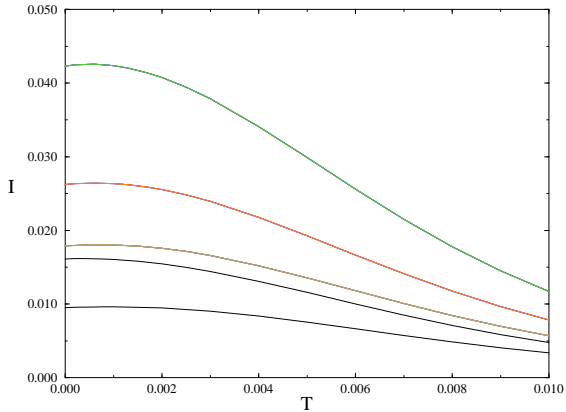, height=7cm, width=7cm,
         bbllx=0, bblly=0, bburx=600, bbury=600,
         angle=270}

\caption{Plots of the critical current (in units of $e v_F k_c$)
versus $T/E_c$ for different strengths of the interaction.
From top to bottom, the solid curves correspond to
$2 e^2/(\pi^2 v_F) = 0.05, 0.2$, and the dotted (colored) lines 
to the latter interaction corrected by the additional short-range
attraction with coupling  $4g/(\pi v_F) = 1.0, 0.75, 0.5$ .}
\label{four}
\end{figure}

We show in Fig. \ref{four} the plots of ${\cal I}_L (T)$, 
including the results obtained by adding the short-range attractive 
interaction. The critical current in the sample $R0_3$
at $T \approx 0 {\rm K}$ is ${\cal I} \approx 2.5 {\rm \mu A}$.
We observe that the correct order of magnitude can be obtained
from our results by considering a renormalization
of the coupling $2 e^2/(\pi^2 v_F)$ down to a value $\approx 0.2$,
together with the effect of a weak short-range attractive
interaction with coupling $g/(\pi v_F) \sim 0.15$.
As a final check of the consistency
of our approach, we observe that the curves in Fig. \ref{four}
reproduce the dependence on temperature measured experimentally
in the sample $R0_3$, with the characteristic inflection point 
and the very slow decay around the critical temperature
at $T \sim 10^{-2} E_c$\cite{kas}. 

We point out that the value $2 e^2/(\pi^2 v_F) \approx 0.2$ is
actually quite close to what is predicted by integrating out the
high-energy electron modes across two orders of magnitude, from
the angstrom scale to the diameter of the rope ($\approx 20 {\rm
nm}$). This follows from the renormalization group approach
worked out in Ref. \onlinecite{eur} for the limit of a very large 
number $N$ of subbands. Moreover, the required value of $g$ 
matches what is expected from the coupling to the elastic modes 
of the nanotube. The short-range effective attraction can be 
estimated from the modulation of the hopping $t' = \partial t/
\partial a \approx 4.2 {\rm eV} {\rm \AA}^{-1} $, the speed of sound
$v_s \approx 2.1 \times 10^4 {\rm m} {\rm s}^{-1}$, and the mass
$M$ of the atoms. This gives $g/v_F \sim t'^2 a^3 /(M v_s^2 v_F)
\sim 0.2$, which is of the same order of magnitude needed in our
fit.

To summarize, we have seen that a model of interacting electrons
in one dimension is able to explain the order of magnitude as
well as the temperature dependence of the critical currents in
both the $ST_1$ and the $R0_3$ samples of Ref. \onlinecite{kas}.
Our description is free of the shortcomings arising from the
conventional picture of the proximity effect, which relates the
value of the critical supercurrent to the gap $\Delta $ and the
normal resistance $R$ through the expression ${\cal I} = \pi
\Delta / (e R)$. Our approach focuses on the strong correlations
in the one-dimensional electron system, explaining in this way
why the experimental data of the supercurrent do not follow in
general the temperature dependence of the gap in the
superconducting contacts.

Our discussion also stresses the relevance of the coupling to
the elastic modes of the nanotube, which reveals itself through
the presence of a short-range attractive electron interaction.
This is also supported by recent experiments on the intrinsic
superconductivity of ropes of nanotubes\cite{sup}. In a sample
like $R0_3$, it can be already observed that the supercurrent
measured experimentally does not vanish near $T_c$, which is at
odds with the conventional picture of the proximity effect but
in accordance with the results of our model.
This enhancement of the superconducting correlations should
deserve further study, in order to understand the experimental
conditions under which the effect of the short-range attraction 
may dominate over the Coulomb repulsion.

Fruitful discussions with S. Bellucci and F. Guinea are gratefully
acknowledged. This work has been partly supported by CICyT (Spain)
and CAM (Madrid, Spain) through grants PB96/0875 and
07N/0045/98.

\end{document}